
\def\cl{\centerline}
\def\ni{\noindent}
{\nopagenumbers

\vskip 1. true cm
\cl {\bf Reversible Work Transition State Theory:}
\vskip 0.2 true cm
\cl {\bf Application to Dissociative Adsorption of Hydrogen}

\vskip 1.0 true cm
\cl {Gregory Mills and Hannes J\'onsson}

\vskip .4 true cm
\cl {\it Department of Chemistry, BG-10}
\cl {\it University of Washington}
\cl {\it Seattle, WA 98195}

\vskip 0.4 true cm
\cl {Gregory K. Schenter}
\vskip .5 true cm
\cl {\it Molecular Science Research Center}
\cl {\it Pacific Northwest Laboratory}
\cl {\it Richland, WA 99352}


\vskip 0.7  true cm
\cl { ({\it Surface Science}, in press)}

\vskip 0.7 true cm
\cl {\bf  Abstract}
\vskip .3 true cm

A practical method for finding free energy barriers for transitions
in high-dimensional classical and quantum
systems is presented and used to calculate the dissociative sticking
probability of $H_2$ on a metal surface
within transition state theory.
The reversible work involved in shifting the system confined to a
hyperplane from the reactant region towards products is evaluated
directly.
Quantum mechanical degrees of freedom are included by using Feynman Path
Integrals with the hyperplane constraint applied to the centroid
of the cyclic paths.
An optimal dividing surface for the rate estimated by transition state theory
is identified naturally in the course of the reversible work evaluation.
The free energy barrier is determined relative to the reactant state
directly so that an estimate of the transition rate can be obtained without
requiring a solvable
reference model for the transition state.
The method has been applied to calculations of the sticking probability
of a thermalized hydrogen gas on a Cu(110) surface.
The two hydrogen atoms and eight surface Cu atoms were included
quantum mechanically and over two hundred atoms in the Cu crystal
where included classically.  The activation energy for adsorption and
desorption was determined and found to be significantly lowered by
tunneling at low temperature.  The calculated values agree quite well
with experimental estimates for adsorption and desorption.
Dynamical corrections to the classical transition state theory rate
estimate were
evaluated and found to be small.

\vfill\eject
}

\pageno 1

\ni \cl {\bf 1. \ Introduction}

The dissociative adsorption of molecules on surfaces of solids
is of central importance in surface catalysis and has been extensively
studied both experimentally and theoretically.
Hydrogen adsorption on copper surfaces has become
`the classic example of activated dissociative adsorption of a molecule at a
surface'.$^{1}$
Many experimental studies
have made use of molecular beams with
molecules in a selected initial state impinging on the surface.
Theoretical studies have focused on classical
trajectory calculations and quantum wavepacket propagation to explore the
dissociative sticking dynamics given a well defined initial state of the
molecule.$^1$
A great deal of qualitative insight has been gained, but
many questions remain open.  On the theoretical side, the problem is that
full quantum mechanical treatment of all
six hydrogen degrees of freedom is impractical with present computers
and currently available techniques for
quantum wavepacket propagation.
The theoretical work has therefore largely been confined to quantum
calculations on lower dimensional model
systems$^{2-4}$ or mixed quantum and classical treatment.$^{5}$
Not only is a six dimensional wavepacket propagation a big order, but
such a calculation would need to be repeated many times to
average over the surface vibrational degrees of freedom to get a single
value of the sticking probability.
Low dimensional simulations have demonstrated clear quantum effects,$^{1,2}$
and calculations in higher dimensions have indicated that
dimensional effects might be equally important.$^{3,4}$

We report here on a fully quantum and thermally averaged simulation
of the dissociative sticking of $H_2$ on $Cu(110)$. The large number of
degrees of freedom can be included at the expense of the amount of
information obtained about the dynamics.
We apply transition state theory, a statistical theory of rates,
to this problem and study entropic and quantum effects on
the sticking probability.
Quantum effects are included by Feynman path integrals (FPIs) which implicitly
include thermal averaging over quantum states.$^{6}$
The simulation mimics a thermodynamic experiment, where the impinging molecules
have a thermal energy distribution at the temperature of the substrate.
This experiment has been carried out in the laboratory of Campbell and
coworkers$^{7}$ who obtained
the thermodynamic activation barrier for dissociative adsorption
from the temperature dependence of the sticking probability.
Our simulations using a rather simple model potential surface
show a clear onset of a quantum mechanical regime
as the activation energy drops at a temperature
around $400 K$.  The experimental
measurements of the sticking coefficient were taken slightly above
this transition temperature.

The method we present here involves evaluation of reversible work
to determine the free energy of the system along the reaction coordinate
and, in particular, the free energy barrier for the transition.
Traditional methods for evaluating free energy differences between two states,
in
particular a transition state and a reactant state involve integration (using
Monte
Carlo sampling) over a parameter $\lambda$ which smoothly converts the
potential
from one state to another.$^{8}$  The advantage of the present technique
is that the transition state need not be known beforehand, and since the
integration
is over the reaction coordinate the intermediate states in the calculation
have some physical significance.
Voter$^{9}$ has presented a technique for direct Monte Carlo calculation of
free
energy differences between separated, localized states by translating one to
overlap
the other.  This method eliminates the need for sampling intermediate
states but again requires a priori knowledge of the transition state,
and can have sampling problems if the transition state potential is shaped much
differently than a slice of the reactant state potential.

The method described here has previously been applied to calculate transition
rates in
a model system consisting of an Eckart barrier coupled to a harmonic
oscillator$^{10}$. Even in such a small system, the reversible work
evaluation was more efficient than calculation of the partition function
with a harmonic reference.
Here we give expressions directly applicable to large systems and
apply the method to hydrogen dissociative adsorption.  Preliminary
results of these calculations have been presented elsewhere.$^{11}$

The organization of this paper is as follows.
We first discuss classical Transition State Theory (TST) in
section 2.  Then quantum TST is discussed in section 3.
In section 4 we present a derivation of the equations required for calculating
free energy barriers in classical systems by evaluation of reversible work.
The application of these equations involves
statistical sampling of forces acting on the system while
constraints are applied.  Following Gillan,$^{12}$ we then generalize these
expressions to quantum systems in section 5
by representing quantum particles by
Feynman path integrals and applying the constraints to the path integral
centroids.
In section 6 we discuss the hydrogen dissociative sticking problem
and present our results on the $H_2-Cu(110)$.
A summary is presented in section 7.

\vskip 1. true cm
\cl {\bf 2. \  Classical Transition State Theory}

Transition state theory is a method for estimating rate constants of
transitions, for example, chemical reactions and diffusion events.$^{13-15}$
It is applicable to slow transitions or rare events.
It is a statistical theory, but is based on dynamics.  The underlying idea is
that the equilibrium density at the reaction's transition state -- a
statistical quantity -- is proportional to the reaction rate -- a dynamical
quantity.
A central aspect of transition state theory is the identification of
a dividing surface
separating  the initial state (the reactants) and the final state (the
products).
Such a dividing surface is frequently referred to as the `transition state'.
All possible transitions from the initial state to the final state
take the system through the dividing surface.

A well-chosen dividing surface also separates the initial and final states
in terms of dynamics.  A system at such a transition state may become either
reactants or products, and once it becomes one or the other, it is likely to
remain so for a much longer time than it spent near the transition state.
Thus, the system's character in the future is strongly dependent on what it
does during its brief visit to the transition state.  This makes the transition
state and what happens there of
central importance.

The system consists of the degrees of freedom of primary interest and all
degrees of freedom to which they are coupled.  For instance, in the $H_2$
dissociation on Cu, the system consists of both the dimer and the Cu lattice.
The hydrogens move furthest and are the primary degrees of freedom, and the
reaction's progress - the reaction coordinate - is determined by the hydrogens'
positions. The surface degrees of freedom are also specifically included
but do not enter the reaction coordinate or the definition of the
transition state; they are effectively a bath.

\vskip 0.8 true cm
\ni {\it 2.1\ The Central Assumptions of TST}
\vskip 0.4 true cm

The first central assumption of transition state theory is that the
reactants have a Boltzmann distribution of
energy in every degree of freedom and that the rate of
transition is low enough that this distribution is maintained.

As a consequence, the system will have a Boltzmann distribution of energy
at the dividing
surface, even if the system doesn't exchange energy with a bath during its
climb up the barrier.  The density at the barrier is low because the barrier
turns most trajectories around.  Those trajectories which do make it to the
dividing surface, however, have a momentum and spatial distribution at the
barrier which is Boltzmann at the reactant temperature except that states
originating
from products are missing.$^{14,16}$
It is important to reallize that TST does not need to make any assumptions
about the ability of the system to achieve an equilibrium distribution at
the dividing surface.

If the reactants are specifically prepared in a non-Boltzmann distribution,
for example, in a molecular-beam experiment where different reactant modes may
be at different temperature, or the surface and adsorbate temperatures are
different, then the system -- consisting of both the incident molecule and the
surface -- is not at thermal equilibrium, and TST does not apply.
We calculate here the sticking probability of a thermalized gas, where
the gas and surface temperature are the same.

The second central assumption of transition state theory is that the
system crosses
the dividing surface only once in going between the reactant and product
states. A reactive trajectory is one which takes the system from steady-state
reactants to steady-state products.
The actual reaction rate depends on the density of reactive {\it trajectories}
at the dividing surface.  Transition state theory assumes that any
crossing of the dividing surface in the direction from reactants to
products is a reactive trajectory.  In reality, trajectories can
cross the dividing surface many times.
Any trajectoriy that crosses the dividing surface an even number of times,
starting from either side, is not reactive, but is counted
in TST.  Further, a trajectory which crosses several times before ultimately
reacting is reactive, but is counted several
times in TST while only forming one product.  As a result, the TST
estimate of the rate is always an overestimate, but in favorable cases and
for a good choice of the dividing surface TST can give an estimate that is
nearly exact.

The fraction of trajectories that have been overcounted in the TST estimate can
be found only by studying
the dynamics of the system.
Transition state theory assumes that the probability, $\kappa$, of continuing
to products, once the dividing surface is reached with a momentum directed
towards
products, is unity.  If the system sometimes recrosses the dividing surface,
$\kappa < 1$.
For a given choice of the dividing surface a
transmission coefficient, $\kappa$, can be found such that
the exact rate constant
is $k_{exact} = \kappa \ k^{TST}.$

For a good choice of the dividing surface, $\kappa$ is as large as possible,
and $k^{TST}$ is as close as possible to the actual rate constant.
A dividing surface far into the reactant state would result in many crossings,
most of which would lack enough energy to cross the barrier and would soon
recross it.  For this dividing surface, $\kappa \ll 1$, and $k^{TST}$ is a poor
estimate of the rate.  Similarly, a dividing surface far into the product state
would be a poor choice, because most of the density there comes not from paths
that have recently reacted, but from products that have recently tried and
failed to back-react.

While TST predicts that the sticking coefficient
depends only on interatomic forces and not on atomic masses,
the recrossings can be mass dependent,
for example, if the products re-desorb because energy is not redistributed
between different modes on the time-scale of the reaction.
This phonon coupling is mass-dependent, and hence can lead to
kinetic isotope effects that are not predicted by TST.

Below, we describe studies we have carried out
of dynamical trajectories for $H_2$
dissociative sticking on $Cu(110)$. We found that dynamical corrections
play a minor role in this system. TST can therefore
be expected to give reliable estimates of the sticking coefficient.

\vskip 1. true cm
\ni {\it 2.2 \ Variational Transition State Theory}

Since classical TST gives an upper bound to the rate, and can be applied to any
dividing
surface between reactants and products, it is possible to search for
the dividing surface for which the estimated rate becomes a minimum; this
estimate is closest to
the true rate.
This dividing surface will have the fewest recrossings.

This optimization of the TST estimate is called Variational Transition State
Theory (VTST)$^{17}$.
The variation of the plane can be done separately at different temperatures, or
at different initial reactant momenta, or by making the dividing surface
non-planar, any added flexibility giving an improved rate estimate.
There exist hypersurfaces that separate the dynamics completely; in general,
these are curved, and different for each reactant energy.$^{13,18}$
If such a hypersurface could be found, the rate associated with it would be the
exact reaction rate:  there would be no recrossings.  However,
for high-dimensional
systems, finding an optimal surface, even within a restricted set (such as
hyperplanar surfaces), is a non-trivial problem.

\ni {\it 2.3 \ The Transition State Partition Function}

Figure 1 shows a reaction path, ${\bf \Gamma}(s)$, between reactants and
products, parameterized with reaction coordinate $s$.  If
$N_p$ is the number of atoms of primary interest; that is, those
that undergo a significant displacement during the transition,
then
${\bf \Gamma}_s$ is a $3N_p$-dimensional vector pointing at position $s$ on the
path.   The reaction path is
defined in terms of these coordinates and is independent of all others.
The dividing surface, $Z^{\ddag}$, is a $~(3N_p-1)~$-dimensional cut through
the
$3N_p$-dimensional system.     In the
case of the $H_2$/Cu system, only the hydrogens change position significantly
during the reaction, so the dividing surface is a 5-dimensional cut through the
6-dimensional $H_2$ system.  All degrees of freedom independent of ${\bf
\Gamma}$ comprise the bath.  In the $H_2$/Cu system, the bath consists of all
the Cu atoms.  If, for instance, two of the Cu atoms moved significantly to
enable the dissociation to happen, the reaction path would include those Cu
atoms, and all other Cu atoms would be the bath.

The dividing surface has one dimension fewer than does the reactant state.
A transition state with the same dimensionality as the reactant can be
defined by arbitrarily introducing a small width $\delta$ to the dividing
surface.
Eventually this width is cancelled out and does not enter the final
expressions.
The probability of finding the system at the transition state is $Q^*/Q^R$.
The transition state partition function
can be written as $Q^* = Q^{\ddag} \delta$, where $Q^{\ddag}$ is the partition
function, taken over the dividing surface, over all system coordinates other
than the reaction coordinate.

If the dividing surface, $Z^{\ddag}$, is specified to be a hyperplane (see fig.
1)
with unit normal vector {\bf n} intersecting the path at ${\bf \Gamma}_{\ddag}$
and letting  {\bf r} be the system coordinate in the hyperplane, then
the partition function is
$$Q^{\ddag} =  \int e^{-\beta V({\bf r})} \delta \left[ {{\bf n} \cdot ({\bf r
- \Gamma}_{\ddag})} \right] d{\bf r} \eqno (1)$$
\ni
where $V$ is the interaction potential, $\beta = 1/{k_B T}$, $k_B$ is the
Boltzmann constant and $T$ is the temperature of the system.  The
Dirac $\delta-$function imposes the hyperplane constraint.
We will deal only with hyperplanar dividing surfaces in this paper.

\vskip 1. true cm
\ni {\it 2.4 \ The Reaction Rate}

Once a system has reached the transition region, its rate of escape is $1/t =
v_{\perp}/\delta$, where $t$ is the time it spends there; $v_{\perp} = {\bf v
\cdot n}$ is its velocity perpendicular to the hyperplane.  To find the
 forward flux, only those points in the transition region that are
forward-bound should be counted.
These comprise half the population there.
The escape rate constant is $Q^*/Q^R$ times the average of $1/t$:
$$k \ =\  {{\langle |v_{\perp}| \rangle} \over 2} \ {{Q^{\ddag}} \over
{Q^R}}.\eqno (2)$$
The factors of $\delta$ in $1/t$ and $Q^{\ddag}$ have cancelled.  Only
positive velocities count because back reactions are ignored.
The average forward velocity is
$${{\langle |v_{\perp}| \rangle} \over 2} = {{k_BT}\over{\sqrt {2\pi\mu
k_BT}}}.\eqno (3)$$
The TST estimate for the rate constant is
$$k^{TST} \ =\  {{k_BT}\over{\sqrt {2\pi\mu k_BT}}}\  {{Q^{\ddag}}\over{Q^R}}.
\eqno (4)$$
Here $\mu$ is an effective mass for the reaction coordinate.

\vskip 1 true cm
\cl {\bf 3. \ Quantum Transition State Theory}

Several different extensions of classical TST to quantum systems
have been proposed.$^{12,19-27}$  We will focus here on
methods where the classical statistical averaging is
replaced by quantum statistical averaging.

\ni {\it 3.1 \ Quantum Statistical Mechanics}

Feynman and Hibbs$^{8}$ describe a reformulation of quantum mechanics in
terms of Feynman path integrals (FPI).  This method involves integrating over
all possible paths a system could take from one place to another.  If done in
real time, it gives quantum dynamics.  In imaginary time, it gives quantum
statistics.  In complex time, it gives a combination of these.

The quantum mechanical partition function, $Q$, can be written as
$$ Q = \int e^{-V_{eff}[x(\tau)]/k_B T} Dx(\tau), \eqno(5) $$
\ni
where $Dx(\tau)$ includes $P$ integrals over coordinates $x(\tau) \approx
\lbrace x_1, x_2, \ldots, x_P \rbrace$ comprising cyclic paths.
The paths are represented to resolution $P$ in this discrete approximation.
Each path contributes to $Q$ according to its effective potential energy,
$V_{eff}$
$^{8,28}$
$$V_{eff} = \sum_{i=1}^P \left[{{k_{spr}(x_i - x_{i-1})^2} \over 2} + {{V(x_i)}
\over P}\right] \eqno (6)$$
with the spring constant given by $k_{spr} = mP/\hbar^2 \beta^2$ where $m$ is
the particle mass and $\beta = 1/k_B T$.  Eqn. (5) becomes exact as
$P \to \infty$.

The classical system has effectively been replaced by $P$ images of itself.
Each image is subject to forces $1/P$ as strong as the classical system
and interacts with its neighbors through the harmonic springs.  In the
classical limit as the mass or temperature become larger, $k_{spr} \rightarrow
\infty$,
and the stiff springs prevent any delocalization.
The $P$ images combine to give back the original classical system.

With $P = 1$ the classical approximation is recovered.
For $P > 1$, the springs allow the chain to delocalize, which gives quantum
statistical effects.  If the chain is at a barrier, it can drape down on either
side to sample a lower effective potential; this represents tunnelling.
If the chain is confined in a small area (as is a particle in a box), it will
have fewer available arrangements,
which gives it a lower entropy.  This confinement entropy corresponds to
zero-point energy.

\vskip 1 true cm
\ni {\it 3.2 \ Centroid Density based QTST}

A rate theory for transitions in quantum systems based on FPI
was introduced by Gillan.$^{12}$
He calculated the thermal equilibrium probability density of the
quantum particle at various locations
by introducing a constraint on the system, fixing
the centroid of the FPI chain
$$ \tilde x_0 \ = \ {1 \over P} \ \sum_i^P \ x_i \ \eqno (7)$$
at a given point,
and taking the average over the remaining quantum degrees of freedom and
classical bath degrees of freedom.
In particular, he evaluated the probability of finding the
centroid at the saddle point of the potential surface
between reactants and products, relative
to the probability of finding the centroid in the reactant region.
By gradually shifting the centroid constraint from the reactant region
up to the saddle point  and monitoring
the reversible work, he obtained a free energy difference.
A rate estimate was obtained by assuming the rate is proportional
to the relative density of the centroid at the saddle point.

In one dimensional systems the classical limit of this estimate
agrees with classical TST.
However, in higher dimensions it does not.
The centroid in Gillan's formulation is confined to a point rather than to a
dividing surface; the transition rate is related to
the centroid density at two specified points.$^{10,12}$
This estimate includes bath degrees of freedom, and various FPI chain
configurations of the quantum particle (thus it picks up some tunnelling and
zero-point energy effects), but it does not include averaging over
centroid degrees of freedom {\it perpendicular} to the reaction path.  Thus, it
does not fully include effects due to varying width of the reaction channel.

Following Gillan's idea, Voth, Chandler, and Miller (VCM)$^{22}$ provided a
quantum generalization of traditional TST by
replacing the classical coordinate in eqn. (1) by the
centroid coordinate of the quantum particle.  Again, a constrained FPI
formalism was used, but with the centroid restricted to a
multidimensional dividing
surface rather than to a point.
The calculations were done by explicit evaluation of the partition function,
using an analytically solvable reference problem.
This method was applied to calculations on a restricted geometry of the
reaction $H + H_2 \to H_2 + H$ and found to give results in good agreement
with more accurate calculations.

Messina, Schenter, and Garrett (MSG)$^{24,25}$ generalized the VCM expression
and allowed for variational optimization of non-planar and momentum-dependent
dividing surfaces.
The dividing surface was chosen to minimize the calculated rate obtained
by direct evaluation of the dividing surface partition function.
MSG is a quantum analogue of classical VTST.
In these calculations a Monte Carlo sampling procedure is used
to find the partition functions explicitly.

These centroid density methods have been applied to various test
problems and have been shown to give good rate estimates.
Transitions in three dimensional systems involving one quantum
particle have also been studied using these techniques.
Gillan studied the diffusion of a hydrogen atom in bulk metals.$^{12}$
Mattsson, Engberg and Wahnstr\"om$^{27}$ and Sun and Voth$^{28}$ studied
the diffusion of a hydrogen atom on Ni(100) and Cu(100) surfaces, respectively.

We describe below a different computational method, reversible work based TST,
for carrying out classical and quantum TST calculations.
For classical systems the results are equivalent to VTST.
For quantum systems, the results are equivalent to MSG.
The advantage of the reversible work formulation is that the method can
more easily be applied to high dimensional systems.  After
describing the method
and its foundation, first as it applies to classical systems (section 4)
and
then as it applies to quantum systems (section 5), we then describe an
application to $H_2$ dissociative adsorption, a reaction path involving
two quantum particles (the H atoms)
and a bath involving eight quantum particles
(surface Cu atoms)
and a couple of hundred classical particles (section 6).
This is by far the largest application of QTST we are aware of.
Our conclusions are given in section 7.

\vskip 1.5 true cm

\cl {\bf 4. \ Reversible Work Formulation of classical TST}

$k^{TST}$ is written in terms of a partition function ratio, or a free energy
difference.  The free energy can be calculated by the method of reversible work
rather than by calculating the partition functions explicitly.
The rate constant, eqn (4),  can also be written

$$k^{TST} = {{k_BT}\over{\sqrt {2\pi\mu k_BT}}} {{Q^{Z^R}} \over {Q^R}}
{{Q^{\ddag}}\over{Q^{Z^R}}} \eqno (8)$$

\ni
where we have introduced $Q^{Z^R}$ a partition function for the system confined
to a reference hyperplane $Z^R$ in the reactant region.  ${{Q^{Z^R}} / {Q^R}}$
has units of inverse length.
${{Q^{\ddag}}\over{Q^{Z^R}}}$ is unitless and, if there are no recrossings of
the transition state, is the reaction probability:  (number of
reactions)/(number of attempts).  The rate is (average positive velocity)
$\times$ (average reciprocal length) $\times$ (reaction probability).  For
surface adsorption, ${{Q^{\ddag}}\over{Q^{Z^R}}}$ is the sticking coefficient.

\vskip 1. true cm
\ni {\it 4.1 \ Relative probability and free energy}

The reaction probability can also be written in terms of the free energy,$A$,
of the transition and reference planes:
$$ \eqalign {
{{Q^{\ddag}} / {Q^{Z^R}}} &= e^{-\beta (A^{\ddag} - A^{Z^R})}, \cr
\Delta A &\equiv A^{\ddag} - A^{Z^R} \cr} \eqno (9)$$
In the reversible work evaluation
the hyperplane is moved stepwise from $Z^R$ to $Z^{\ddag}$, as indicated in
figure 2, and the free energy difference, $\Delta A$, is accumulated at each
step.

Figure 1b shows a progression of hyperplanes $Z_s$ defined along the entire
reaction path ${\bf \Gamma}_s$ between reactants and products.
Given a path, the
sequence of hyperplanes can be constructed by choosing the tangent vector
of the path
${d \over {ds}}{\bf \Gamma}_s$,
at various points along the path as the normal to a
hyperplane associated with that reaction coordinate
$${\bf n}_s \equiv {d \over {ds}} {\bf \Gamma_s} \ / \ \left| {d \over {ds}}
{\bf \Gamma_s}
\right|. \eqno (10)$$
The progression of the hyperplane has to be gradual enough that a first order
evaluation of the free energy difference will be accurate.
As the reaction coordinate is incremented by $ds$, where
$$ds \ = \|{\bf \Gamma}_{s+ds} - {\bf \Gamma}_s\|,\eqno (11)$$
the hyperplane intersection with the path moves from
${\bf \Gamma}_s$ to ${\bf \Gamma}_{s+ds}$ and the
normal vector changes from ${\bf n}_s$ to ${\bf n}_{s+ds}$.

It is convenient to introduce a coordinate system aligned with the
hyperplane.
These local coordinates, ${\bf z}$, are obtained from the actual coordinates,
${\bf r}$, by applying a unitary rotation matrix $U$:
$$\eqalign {{\bf r}_s - {\bf \Gamma}_s &= U_s {\bf z}, \cr
{\bf r}_{s+ds} - {\bf \Gamma}_{s+ds} &= U_{s+ds} {\bf z}. \cr}\eqno (12)$$
\ni
We choose the first unit vector in the local coordinate system to
be normal to the hyperplane.
The hyperplane constraint, $({\bf r}_s - {\bf \Gamma}_s) \cdot {\bf n}_s = 0$,
can then be written as $z_1 = 0$.  The first column of $U_s$ is ${\bf n}_s$.

The free energy difference corresponding to an increment $ds$
in the reaction coordinate is $dA = -k_BT \ln (Q_{s+ds}/Q_s)$ where
$$\eqalign {
Q_s &= \int e^{-\beta V_s({\bf z})} \delta (z_1) d{\bf z} \cr
Q_{s+ds} &= \int e^{-\beta V_{s+ds}({\bf z})} \delta (z_1) d{\bf z}. \cr}\eqno
(13)$$
\ni
{}From the point of view of a system confined within the hyperplane,
the potential energy changes as the reaction coordinate is incremented
$$\eqalign {
V_s({\bf z}) &\equiv V({\bf r}_s) \cr
V_{s+ds}({\bf z}) &\equiv V({\bf r}_{s+ds}) . \cr}\eqno (14)$$
\ni
This ramping of the potential energy of the system is illustrated in fig. 2.
Since the planes are nearby,
$$V_{s+ds}({\bf z}) = V_s({\bf z}) + \delta V({\bf z}),\eqno (15)$$
\ni
where $\delta V({\bf z})$ is small for all values of ${\bf z}$
the system is likely to visit.

The partition function ratio is
$${{Q_{s+ds}}\over{Q_s}} = {{\int e^{-\beta \delta V({\bf z})} e^{-\beta
V_s({\bf z})} \delta (z_1) d{\bf z}}\over{\int e^{-\beta V_s({\bf z})} \delta
(z_1) d{\bf z}}}.\eqno (16)$$
\ni
which can be written as

$${{Q_{s+ds}}\over{Q_s}} = \langle e^{-\beta \delta V({\bf z})}\rangle_s
= e^{-\beta  \langle \delta V({\bf z})\rangle_s} + O\left(\delta V ^2\right)
\eqno (17)$$
\ni
since $\delta V$ is small.  The average is calculated in plane $Z_s$.
Thus, if the partition functions are written in terms of the constant internal
coordinate ${\bf z}$,  moving the hyperplane from one position
to another is, in effect, the same as ramping the potential energy seen by
a system confined to the hyperplane by an amount $\delta V({\bf z}).$
The free energy change becomes simply the thermal average of the local internal
potential ramp $\delta V({\bf z})$:
$$dA = \langle\delta V({\bf z})\rangle_s + O\left(\delta V ^2\right).
\eqno (18)$$

The external coordinates of the system in the two hyperplane locations
are related by
$${\bf r}_{s+ds} - {\bf \Gamma}_{s+ds} = U_{s+ds} U_s^T ({\bf r}_s - {\bf
\Gamma}_s).\eqno (19)$$
\ni
Writing $U_{s+ds} = U_s + \delta U_s$, $U_{s+ds} U_s^T = U_s U_s^T + \delta U_s
U_s^T = I + \delta U_s U_s^T$  and substituting into the equation above
gives, after rearranging,
$$\eqalign {{\bf r}_{s+ds} - {\bf r}_s &= ({\bf \Gamma}_{s+ds} - {\bf
\Gamma}_s) + \delta U_s U_s^T ({\bf r}_s - {\bf \Gamma}_s) \cr
&= {\bf n}_s ds + \delta U_s U_s^T ({\bf r}_s - {\bf \Gamma}_s) \cr}\eqno
(20)$$
using the fact that we chose the normal to be tangent to the path, eqn (10).
The first column of $U_s$ is ${\bf n}_s$.  The other columns can be assigned
arbitrarily; the free energy of the plane is independent of them.  These column
vectors ${\bf u}_i$ comprise an orthonormal basis for the hyperplane.  The
geometrical difference between the two locations of the hyperplane is
entirely specified by ${\bf \Gamma}_s$, ${\bf \Gamma}_{s+ds}$, ${\bf n}_s$, and
${\bf n}_{s+ds}$.  We will now show that the free energy change can be
expressed
entirely in terms of this geometry, without explicit calculation of the
internal basis vectors.

If $D$ is the dimensionality of the primary system ($3N_p$ in a three
dimensional system), then
$$ \delta U_s = \left\lbrack \matrix {
d{\bf n}_s & d{\bf u}_2 & \ldots & d{\bf u}_D \cr}
\right\rbrack. \eqno (21)$$

\ni
Since ${\bf u}_i \cdot {\bf u}_j = \delta_{ij}$ for an orthonormal basis, ${d \
\over {ds}} ({\bf u}_i \cdot {\bf u}_j) = 0$.  This property makes the matrix
$\delta U_s U_s^T$ antisymmetric.

$$ \delta U_s U_s^T = \left\lbrack \matrix {
d{\bf n}_s & d{\bf u}_2 & \ldots & d{\bf u}_D \cr}
\right\rbrack
\times
\left\lbrack \matrix {
{\bf n}_s \cr
{\bf u}_2 \cr
\vdots    \cr
{\bf u}_D \cr}
\right\rbrack. \eqno (22)$$

\ni
Writing, for convenience, ${\bf n}_s$ as ${\bf u}_1$, and making use of the
antisymmetry, this can be written as

$$ \delta U_s U_s^T = \left\lbrack \matrix {
0                      & -\sum du_{i2} u_{i1} & \ldots & -\sum du_{iD} u_{i1}
\cr
-\sum du_{i1} u_{i2} & 0                      & \ldots & -\sum du_{iD} u_{i2}
\cr
\vdots                 & \vdots                 & \ddots & \vdots
  \cr
-\sum du_{i1} u_{iD} & -\sum du_{i2} u_{iD} & \ldots & 0
\cr}
\right\rbrack, \eqno (23)$$

\ni
where $u_{ij}$ is the $j$th component of the $i$th basis vector.

Operating with this matrix on the vector ${\bf R} \equiv {\bf r}_s - {\bf
\Gamma}_s$ yields, after rearrangement,

$$ \delta U_s U_s^T ({\bf R}) = - \sum_i {\bf u}_i (d{\bf u}_i \cdot {\bf
R}).\eqno (24)$$

The internal potential energy ramp is

$$ \eqalign {
\delta V({\bf z}) &= V({\bf r}_{s+ds}) - V({\bf r}_s) \
= -{\bf F}({\bf r}_s) \cdot ({\bf r}_{s+ds} - {\bf r}_s) \cr
&= -{\bf F}({\bf r}_s) \cdot {\bf n}_s ds - \
{\bf F}({\bf r}_s) \cdot (\delta U_s U_s^T ({\bf R})) \cr
&= -{\bf F}({\bf r}_s) \cdot {\bf n}_s (ds - d{\bf n}_s \cdot {\bf R}) + \
\sum_{i=2}^D ({\bf F} \cdot {\bf u}_i) (d{\bf u}_i \cdot {\bf R}). \cr} \eqno
(25)$$

The thermal average of each term in the sum over the in-plane basis vectors is
zero,
$\langle ({\bf F} \cdot {\bf u}_i) (d{\bf u}_i \cdot {\bf R}) \rangle = 0$.
Since the unit basis vector ${\bf u}_i$ is orthogonal to its change $d{\bf
u}_i$, the average of the product is the product of the averages.  $\langle
({\bf F} \cdot {\bf u}_i) \rangle = 0$ for any in-plane ${\bf u}_i$, whether or
not any orthogonal coordinates are specified.  $(d{\bf u}_i \cdot {\bf R})
\langle {\bf F} \cdot {\bf u}_i \rangle$ is therefore individually zero for
each specified $(d{\bf u}_i \cdot {\bf R})$, and the overall average vanishes.

The angle between the planes' normal vectors, ${\bf n}_s$ and ${\bf n}_{s+ds}$,
is $d\theta = \| d{\bf n}_s \|$.  The first-order curvature of the reaction
path ${\bf \Gamma}$ at $s$ is $\kappa = d\theta / ds$.  We write for the normal
component of the force, $F_n \equiv {\bf F}({\bf r}_s) \cdot {\bf n}_s$, and
for the turning-direction component of the distance from the reference point,
$R_t \equiv ({\bf r}_s - {\bf \Gamma}_s) \cdot d{\bf n}_s / d\theta$.  Then we
can write the free energy difference between the planes as

$$ dA(s)\  =\  - \ \langle F_n \ (1 - \kappa R_t) \rangle_s \ ds\ . \eqno (26)
$$

\ni
Here $-\langle F_n R_t \rangle_s$ is the average torque opposing the rotation
of the plane around the reference point ${\bf \Gamma}_s$ in the direction
$d{\bf n}_s / d\theta$.  The amount of work required to rotate the plane by
$d\theta$ in this direction is

$$ dA_{rot}(s) \ =\  -\  \langle F_n R_t \rangle_s \ d\theta\ =\  -\  \kappa
\langle F_n R_t \rangle_s\ ds . \eqno (27)$$

The free energy difference between the reference hyperplane, $Z^R$, and a
hyperplane $Z_s$ corresponding to reaction coordinate $s$ is
$$\Delta A(s) \ =\  -\ \int_R^s \ \langle F_n \ (1 - \kappa R_t)
\rangle_{s^\prime } \ d{s^\prime }. \eqno (28)$$

If all coordinates are specified, the instantaneous force is the mean force.
The line integral of the instantaneous force is the potential energy.
The reversible work involved in shifting the system
from one {\it completely specified} configuration to another is just the
potential energy difference.

\vskip 0.5 true cm

\ni {\it 4.2 \ Implementation of the method}

First a reaction path leading from the reactant to the product region is
defined.  A sequence of hyperplanes intersecting the path is then defined.
The hyperplanes must be spaced closely enough for the first order
expression, eqn. (26), to be accurate enough.
The average force and torque acting on the system confined to the hyperplane
are then statistically sampled and the results integrated according to eqn.
(28).

A convenient choice for the path is the minimum energy path (MEP),
but any path that does not
lead to large numerical cancellations in the force evaluation
can be used.
The hyperplane normals can conveniently be chosen to be tangent
to the path (as we have done here), but this is not necessary.
To collect the average force acting on the system in each of the
hyperplanes, a Monte Carlo or a molecular dynamics simulation can
be carried out subject to the constraint
that the system remains in the hyperplane.
More specifically,
considering a primary system consisting of $N_p$ atoms in 3-dimensions,
a $3N_p$-dimensional unit vector,  the hyperplane normal ${\bf n_s}$,
is used to constrain the dynamics.
At each time step the component of the force vector normal to the hyperplane,
${\bf F} \cdot {\bf n_s} $, is collected for the evaluation of the thermal
average, $<{ F_n}>$. But then, this component is
projected out of the force to give a new, revised force
${\bf F}^{rev}  \ = \ {\bf F} \ - \ {\bf F} \cdot {\bf n_s} $
which is used in the simulation giving the statistical sampling.

We have used the velocity Verlet algorithm to
carry out the molecular dynamics sampling of canonical phase space.$^{31}$
The constrained dynamics conserve energy as
long as the constraint vector is constant.
The optimal dividing surface $Z^{\ddag}$ for the TST rate estimate is the
hyperplane $Z_s$
corresponding to the maximum free energy, where force $<{ F_n}>_s $
and the torque $<{ F_n R_t}>_s $ acting on the hyperplane both vanish.
It could be necessary to adjust
the orientation of the hyperplane at and near the transition state in
order to zero the torque.  Alternatively, the path can be redefined
to construct a `zero torque path' as we have done for an Eckart barrier
test problem.$^{10}$
As we will show below, the torque acting on the MEP-constructed hyperplane
in the hydrogen dissociative adsorption simulation
turns out to be very small, leading to negligible rotational
contribution to the free energy barrier.

Each plane has a certain free energy, $\Delta A(s)$, relative to the reference
plane in the reactant state.  This free energy gives a TST estimate for the
rate, treating that plane as the dividing surface:

$$k_s^{TST} = {{k_BT}\over{\sqrt {2\pi\mu k_BT}}} {{Q_s}\over{Q^R}} =
{{k_BT}\over{\sqrt {2\pi\mu k_BT}}} {{Q^R}\over{Q^{Z^R}}} e^{-\beta \Delta
A(s)}. \eqno (29)$$

The plane for which this estimated rate is lowest is the planar dividing
surface at which there are the fewest recrossings.  So VTST is accomplished
automatically by the reversible work calculation.  At some plane crossing the
reaction path, the free energy will be a maximum; this gives the VTST rate
estimate.  The automatic inclusion of VTST is an advantage of the reversible
work method.

The free energy barrier found this way is path-independent as long as the
planes near the barrier are good dividing surfaces.  The optimal planar
dividing surface is the one corresponding to maximum free energy, and therefore
has zero average force and torque acting on it.  If there is a net torque on
the best plane of the sequence, it can be rotated against the torque, picking
up some additional free energy, until the torque is zero.

The idea of using a one-parameter sequence of generalized dividing
surfaces and minimizing the calculated rate constant with
respect to the parameter has been implemented by
Garrett and Truhlar in a semiclassical version of variational transition
state theory.$^{17}$  Although their methods
employ different calculations of the rate than the current work, the idea of
optimizing the rate along a path is equivalent to the determination of
the hyperplane giving the maximum free energy in the present method.

\vskip 1.5 true cm

\cl {\bf  5. \ Reversible work formulation of quantum TST}

\vskip 0.5 true cm

The formulas derived for classical TST will now be generalized to quantum
systems by using quantum statistical mechanics instead of classical statistical
mechanics to estimate the free energy barrier.  The central
question is how to incorporate the constraints on the system;
i.e. how to confine the system to a hyperplane.
Following Gillan$^{12}$ we represent each quantum particle in the
system with a Feynman Path Integral and apply the constraints to the
centroids of the path integrals instead of the atom coordinates as
in the classical case.

The hyperplane constraint then becomes
$$Z_s = \  (\tilde {\bf r}_0 - {\bf \Gamma}_s) \cdot {\bf n}_s = 0, \eqno
(30)$$
where $ \tilde {\bf r}_0 $ is a vector containing the controid coordinates
of quantum atom FPIs.  Using a discrete representation of the paths,
the centroid for atom $k$ is
$${\bf \tilde r}^k_0 \ =\ {1 \over P} \
\sum_{i=1}^P \ {\bf r}^k(i) \eqno (31)$$
where the sum extends over all the FPI images of atom $k$.
The reaction's progress is written in terms of the FPI centroids of the
quantum mechanical atoms in the primary system.  Therefore, the same
reaction path can be used as in the classical calculation.

The quantum mechanical partition function for a hyperplane at $s$ is:
$$Q_s\  =\  \int e^{-\beta V_{eff}} \delta \left[ {{\bf n}_s\cdot ({\bf \tilde
r}_0-{\bf \Gamma}_s)} \right] D{\bf r}(\tau ). \eqno (32)$$
with $V_{eff}$ given by eqn. (6).
Just as in the classical case, the change in the hyperplane free energy
$\Delta A(s)$ can be evaluated.
The quantum mechanical generalization of eq. (28) is
$$\Delta A(s) \ =\  -\ \int_R^s \ \langle \
\ {1 \over P} \  \sum_{i=1}^P \
F_n(i) \ (1 - \kappa R_t(i)) \rangle_{s^\prime}
\ d{ s^\prime} \eqno (33)$$
where the hyperplane statistical average is
$$\left\langle \cdots  \right\rangle _s = {{\int {D{\bf r}(\tau )\
e^{-\beta V_{eff}}\ }\cdots
\ \delta \left[ {{\bf n}_s\cdot \left( {{\bf \tilde r_0 -\Gamma}_s} \right)}
\right]} \over {\int {D{\bf r}(\tau)\
e^{-\beta V_{eff}\ }\delta \left[ {{\bf n}_s\cdot \left({{\bf \tilde r_0
-\Gamma}_s} \right)} \right]}} } . \eqno
(34)$$
In the Appendix we prove that the in-plane components of the force
on the images do not contribute to $\Delta A(s)$, just as in the
classical case.

The integration along the reaction path gives a free energy maximum
and thereby the free energy barrier for the transition in the
quantum system.
The transition rate is then obtained from the free energy barrier
as in the classical case.
Again, the reversible work calculation automatically includes variational
maximization of the hypersurface free energy.  The transition
rate obtained by this reversible work method is exactly the same as the
variationally optimized rate for a hyperplanar dividing
surface obtained in the MSG theory,$^{24,25}$
but the advantage of the present method is
that a solvable reference system for the transition state is not required.

There is no rigorous justification for this formulation of QTST.
Other formulations of a quantum generalization of TST have been
proposed.  The relationship between the estimate obtained from the
QTST presented here and the exact reactive flux$^{32}$ is unclear.
The rate constant for a quantum system calculated by this and other
centroid methods has not been proven to give an upper bound to the rate in
general.
It has been
found that optimization of the dividing surface leads to a strict
upper bound on the rate in the harmonic limit
as well as in a symmetric Eckart barrier test problem.$^{10,24}$
However, $k^{QTST}$ has in some other cases been found smaller (but only
slightly smaller) than $k_{exact}$.

\vskip 1 true cm

\cl {\bf 6.  \ Dissociative Sticking of Hydrogen on Surfaces}

We now turn to the application of the reversible work TST method
to the dissociative sticking problem.
The calculation of the sticking coefficient described here include all six
$H_2$ degrees of freedom quantum mechanically as well as eight Cu surface
atoms
and full thermal averaging over several hundred substrate degrees of freedom.
However, the calculations are done within TST and therefore apply
to sticking of thermalized hydrogen gas, not molecular beam
experiments with selected initial states of the molecules.
The large number of degrees
of freedom can be included at the expense of detailed dynamical information
about the system.

Campbell, Domagala, and Campbell$^{9}$ measured the sticking coefficient of
thermally equilibrated hydrogen gas on $Cu(110)$ experimentally. These
measurements give an activation energy of $0.62~ eV$ in the range
$500~ K$ to $700~ K$.
Molecular beam experiments convolved to yield a thermal distribution
give similar results.$^{33}$
 Also, the activation energy for thermal desorption of $H_2$ from $Cu(110)$ has
been measured by Anger, Winkler, and Rendulic.$^{34}$  The results of our
simulation are compared to these experimental measurements below.

\vskip 1. true cm
\ni {\it 6.1 \   The `hole model' for sticking}

While TST is most often applied to transitions involving large
energy barriers, the foundation of TST
requires only that the rate of transition is slow enough to allow the remaining
reactants to stay at thermal equilibrium; and that once the dividing surface
has been crossed, it will not be recrossed on the time scale
of the transition.
A bottleneck in phase space involving only an entropic barrier can
satisfy these conditions -- an energy barrier is not necessary.

As an example, the TST
assumptions apply well to an ideal gas in a box effusing through a small hole.
 There is no energy barrier; just an entropy barrier. Most atoms colliding with
the wall will fail to escape because the hole is much smaller than the box's
cross-section.
If the box volume is $V$ and the hole has cross sectional area $A$, then
$Q^{\ddag}/Q^R = A/V$
and the TST estimate for the effusion rate is
$$Nk^{TST} = {{A \rho k_BT}\over{\sqrt {2\pi\mu k_BT}}} = {{AP}\over{\sqrt
{2\pi\mu k_BT}}}. \eqno (35)$$
which is precisely the rate given by kinetic theory.

A successful model of
molecular adsorption, the  `hole model',$^{35,36}$  is analogous to the
effusion
problem.  The assumption there is that a molecule approaching the surface
with the right orientation will adsorb.  However, most orientations of the
molecule are unfavorable and lead to a large potential barrier
which reflects the molecule back into the gas phase.
The transition rate is then estimated as the ratio of the volume of the
hole relative to the total volume available to the incoming molecule.
If the hole model is a reasonable model for dissociative sticking,
then TST can be expected to give a good estimate of the sticking
probability.

\ni {\it 6.2 \ The $H_2-Cu(110)$ Potential Energy Surface}

A simple form for the potential function
was used to speed up the calculation.  An EAM-type$^{37}$ potential
form was chosen and fitted to data for the $H_2$ dimer and the chemisorbed
hydrogen.  It was also fitted to a grid of points taken from a semi-empirical
LEPS potential$^{38}$, which is expected to have the
right shape qualitatively, as it includes overlap and exchange effects.
However, the potential barrier was adjusted to have a much
higher saddle point than in ref. 38.
The potential barrier was scaled up to $0.688 \ eV$, close to
the estimate of Hand and Harris based on $ab\  initio$ calculations.$^{39}$
(See table 1).

The EAM form works well for copper. With a filled $d$ band, $Cu$ exhibits
little directional bonding.  Most of the electron density seen by a
neighbor is due to the 4s electron.

The EAM potential form is:
$$V = \sum_{ij} \phi (r_{ij}) + \sum_i F(\rho_i), \eqno (36)$$
\ni
where $\phi$ is a pair potential that represents the screened coulomb repulsion
between the ion cores, including the inner electrons, and F($\rho$) is the
energy of embedding an ion core into a valence electron gas
of density $\rho$.  The electron density $\rho_i$ at atom i is the sum of the
contributions of all its neighbors, j:

$$\rho_i = \sum_j \rho^A(r_{ij}), \eqno (37)$$

\ni
where $\rho^A(r_{ij})$ is fitted to be the Hartree-Fock atomic electron
density.

We have chosen the following functional forms
$$\eqalign {
\phi (r_{ij}) &= D_A e^{-\alpha_A r_{ij}} + D_B e^{-\alpha_B r_{ij}} \cr
F (\rho_i) &= \sum_{m=1}^M f_m \rho_i^m \cr
\rho^A (r_{ij}) &= S r^{\eta} \left( e^{-\beta_A r_{ij}} + \
 \gamma e^{-\beta_B r_{ij}} \right) \cr
}\eqno (38)
$$
$\phi (r_{ij})$ is a pair potential that represents the screened coulomb
repulsion between the ion cores.  It is parametrized as a double exponential.
Each type of pair interaction (Cu-Cu, H-H, and H-Cu) has its own parameter set
$\lbrace D_A, \alpha_A, D_B, \alpha_B \rbrace$.
The parameters for the pair potential,
the embedding energy and the electron density
are given in table 2.

The pair potential and the electron density are shifted in such a way
that they become  zero at and beyond a cutoff distance, $R_{cut}$,
chosen to be $6.1 {\AA}$.

\vskip 1. true cm
\ni {\it  6.3 \ Finding Minimum Energy Path using Nudged Elastic Band method:}

The MEP is defined by the boundary conditions, and by the requirement that the
force at any point be parallel to it.  That is, for any direction ${\bf p}_s$
perpendicular to the path, \ ${\bf p}_s \cdot {d \over {ds}}{\bf \Gamma}_s =
0$, the potential energy is at a minimum:  ${\bf p}_s \cdot {\bf \nabla} V
({\bf \Gamma}_s) = 0$.

The boundary conditions are that the MEP must start at the lowest-energy
reactant state and finish in the lowest-energy product state.  For the $H_2$/Cu
problem, the former has the dimer distant from the surface at its equilibrium
separation;
the latter has the hydrogens on neighboring sites in the same groove on the
(110) surface.

The `Nudged Elastic Band' method$^{40}$ was used to find the path.  This method
finds the MEP variationally.  An elastic band, made of beads connected by
harmonic springs, is strung between the reactant and product states.
Using a set of images or replicas of the system to define a discrete
transition path,
the problem is turned into a minimization problem by defining an object
function
$$ F({\bf r}_1, {\bf r}_2, \dots ,  {\bf r}_{P-1}) \ = \ \sum^{P-1}_{i=1} \
V({\bf  r}_i)
\ + \ \sum^{P}_{i=1} \ {{kP} \over 2} \ ({\bf  r}_i - {\bf r}_{i-1})^2\  \eqno
(39)$$
where the sum is over the `true' potential of all the intermediate
images of the system,
and the second sum is an `elastic glue' or `spring energy'
that keeps adjacent images together
The end point images $\bf r_0$ and ${\bf r}_P$ are fixed.
Although this chain is in some ways
analogous to the off-diagonal density matrix FPI, the MEP is found with no
reference to quantum mechanics.  Here, the spring constant is arbitrary
except that it should scale as $ k P$ to ensure convergence to the
MEP as more images are introduced in the chain, $P \to \infty$.$^{27}$

In order to accelarate the convergence of the chain to the MEP we
introduce `nudging'.
The component of the net spring force on a bead
{\it parallel} to the local path tend to keep the bead equidistant from its
neighbors; whereas the {\it perpendicular} spring force
tends to make it collinear with its neighbors.
When the MEP is curved, the chain will tend
 to pull the chain off the MEP by the perpendicular spring force
(thereby inducing corner cutting).
We therefore zero the perpendicular component of the spring force.
Each bead also feels a force due to the gradient of the `real' potential
energy.  The component of this gradient force {\it perpendicular}
to the path tries to pull the beads down toward the MEP; whereas its {\it
parallel} component tries to pull the beads down toward either the reactant or
product state, away from the saddle point.
This effect would result in reduced resolution
at the saddle point.  We therefore zero the parallel component of this
gradient force.
Moving each bead in the direction of its adjusted force until the adjusted
force becomes zero will put the beads on the MEP (to within the chain's
resolution), equally spaced in the $3N_p$-dimensional space.
The technique is not a global search:  if there is more than one MEP, only one
will be found.

Figure 4 shows a MEP for the $H_2$/Cu problem.  The initial state is asymmetric
with the final state.  Even so, the path becomes symmetric in the vicinity of
the surface, lining up with the groove on the (110) surface.
The Nudged Elastic Band method has therefore shown that the MEP is symmetric.
In a six dimensional system, this is a non-trivial result.

Because of the symmetry found for the MEP, a contour plot of the potential
surface characterizes the path quite well.
Constraining the $H-H$ axis to be parallel to the surface and aligned with a
surface groove, and fixing the $H_2$ center of mass above a long-bridge site,
the two remaining variables are
Z, the $H_2$ dimer's height above the surface, and  R,  the H-H separation.
Figure 5 shows a contour plot of the $H_2$/Cu potential energy surface,
$V(Z,R)$, with the $H_2$ geometry restricted in this way.    The surface
configuration is fixed at the minimum energy for a bare surface.
The energies are in eV relative to the bare surface and two separated hydrogen
atoms.

\vskip 1. true cm
\ni {\it 6.4 \  Statistical sampling of the hyperplanes}

The hyperplane constraints derived from the MEP become quite simple
due to the symmetry.
Taking
the x axis to be along the rows of atoms on the (110) surface,
the y axis across the rows, and the z axis normal to the surface,
the hyperplane constraint derived from the MEP becomes
$\tilde z_1 + \tilde z_2 \ = \ constant$ in the reactant state
and $\tilde x_1 - \tilde x_2\ = \ constant$
in the product state.
Here
$\tilde x_1$ is the x coordinate of hydrogen 1's centroid, etc.
In the intermediate hyperplanes the constraint is
$$a(\tilde x_1 - \tilde x_2) + b(\tilde z_1 + \tilde z_2) = 0. \eqno (40)$$
where
$a$ and $b$ are constants.  This means that within the transition state
hyperplane, for example,
the centroids are allowed to move closer together only if they simultaneously
descend
toward the surface.  They are allowed to move apart only if they ascend higher
above the surface.  This constraint prevents the atoms from either chemisorbing
on adjacent lattice sites, or desorbing.
The plane constraint does not affect symmetric motion along the groove (a
change in $\tilde x_1 + \tilde x_2$), asymmetric vertical motion ($\tilde z_1 -
\tilde z_2$) or the y coordinates perpendicular to the groove.  The plane also
does not prevent individual images from desorbing or adsorbing as much as the
springs between them will allow.

Figure 6 shows scatterplots of the $H-H$ orientations in the classical
statistical sampling at $T=100~K$.
Plotted are the angles theta and phi of the H-H orientation, relative to the
surface.  Theta is the polar angle, between 0 and $\pi$.  Phi is the azimuthal
angle, between $-\pi$ and $\pi$.
Figure 6a is for a hyperplane in the reactant region.  There is no
restriction on the orientation of the molecule from the hyperplane
constraint.
There are fewer points at low and high values of theta because most of the
surface area of a sphere is found near the equator, at theta $\approx \pi/2$.
The scatterplot, when drawn as a rectangle, is therefore not uniform.
In figure 6b is taken at the the transition state hyperplane
where the rotation is highly constricted by the EAM interaction potential.
This orientational confinement of the molecule in the transition state
contributes to a significant entropy barrier for the transition.

In the quantum statistical sampling, the two $H$ atoms and
eight $Cu$ atoms
nearest to the $H$ atoms are included as FPI chains.
We have used $P=50$ images in each FPI chain in
the present calculations.
Figure 7 shows three snapshots of the FPI chains from the
quantum statistical sampling of the transition state hyperplane at $T= 100~ K$.
The system consists of two hydrogen atoms in the same groove of the $Cu(110)$
surface, across the long bridge.  The great flexibility available to
the images in the H atom FPIs is evident.
For illustration purposes only
20 beads where included in the FPI chains in this figure.

The hyperplane constraint is a very loose constraint, allowing the system to
sample a wide range of configurations.
The system slides back and forth in the groove as illustrated in fig. 7.
At low temperatures, where tunneling is important,
the images drape down significantly on
either side of the barrier, as illustrated by the
separation of the images corresponding to the same FPI into two groups in
fig 7.  Few images are actually at the barrier top.
At a higher temperature, the springs are tighter and the distribution of images
remains nearly spherical even at the barrier top.

\vskip 1. true cm
\ni {\it 6.5 \ The free energy barriers}

The accumulated free energy difference with respect to the reactant state
as the hyperplane is moved along the reaction path in
a classical statistical sampling is shown in fig. 8.
The free energy rises faster than the potential energy
along the reaction path.
The free energy barrier increases with temperature, indicating a
drop in entropy as the system approaches the transition state.
The intercept of an
Arrhenius line from several temperatures gives $\Delta S = -3.4 ~ k_B$.

This drop in entropy at the transition state
is a result of the confinement of the molecule (see fig. 6)
by the interaction with the surface as well as confinement of the
lattice vibrations due to the presence of the molecule.
The two effects were studied separately by first holding the lattice
rigid and sampling only the degrees of freedom of the hydrogen
molecule (fig. 9a); and then holding the hydrogen molecule fixed on
the MEP while sampling statistically over the lattice degrees of
freedom (fig. 9b).
In the second case the rotational-vibrational entropy in the
5 in-plane degrees of freedom
of the hydrogen molecule has been removed.
The free energy becomes much closer to the MEP potential and has little
temperature dependence, corresponding to $\Delta S = - 1.2~ k_B$.
The effect of the molecular confinement with the lattice rigid is
much larger, corrsponding to $\Delta S = -5.3 ~ k_B$.
The two effects are not additive. The dynamics and relaxation
(amounting to $ca. \ 0.15~ {\AA}$  displacement of the surface Cu atoms)
reduce the confinement
of the $H_2$ molecule in the transition state, thereby lowering the net
entropy barrier for the dissociative sticking.
In other words, the movable lattice restricts the hydrogen dimer by a smaller
amount than does the rigid lattice.
The lattice dynamics therefore assist the reaction by lowering the entropic
barrier.

The results of the quantum statistical sampling are shown in
Figure 10 for several different temperatures.
The low temperature curves are flat-topped near the transition state.  This is
a tunnelling effect; it is a consequence of the chain's delocalization.
The images drape down over the barrier and manage to avoid the barrier top.
Figure 11 compares the free energy curves for quantum $H_2$, quantum $D_2$, and
the
classical $H_2$, at 100 K.  The deuterium curve shows clear tunneling
effects, but smaller than hydrogen, as expected.
The accumulated free energy in the quantum statistical sampling
also drops below the classical free energy even before the tunnelling region is
reached.
This is evidently a zero-point energy effect due to the softening of the $H-H$
vibration.

We have used a total of 45 hyperplanes to evaluate the free energy barrier.
In the first hyperplane the center of the molecule is $7~ {\AA}$ above the
surface.  The spacing between the hyperplanes was not even
as can be seen from the data points in fig. 8. Initially the spacing is
large since the force is small far from the $Cu$ surface.
The accuracy of this
discretization of the hyperplane progression and the numerical integration of
the force was tested by recalculating the free energy barrier using only
every other hyperplane.  At 100~K, the integrated free energy barrier differed
by $0.0002~ eV$ in the classical simulation and by $0.016 ~ eV$ in
the quantum simulation.
The error due to finite hyperplane spacing is largest in the low-temperature
quantum calculation where tunnelling is most significant, because the free
energy curve is
flat-topped and has a rapidly changing derivative at the onset of tunnelling.

A larger source of error in the free energy calculation are statistical
errors due to the limited simulation time in calculating the average force.
This error is greatest in the high-temperature quantum simulations where the
timestep needs to be very small.  From the standard deviation in the force
acting
on the hyperplanes, we estimate the standard deviation in the accumulated
value of the free energy barrier to be $0.03 ~eV$ in the quantum results at
$600~ K$ - which is the case with largest statistical fluctuations.

\vskip 1. true cm
\ni {\it 6.6 \ Tunneling}

In an effort to separate the tunneling effect from the effect of
changes in zero-point energy,
the RMS delocalization perpendicular to the hyperplane was evaluated.
The RMS delocalization of a free atom in one dimension, as determined by
recursive Gaussian integration$^{8}$ of the quantum partition function
and taking the limit as $P \to \infty$, is
$$\sqrt {\langle (x - \tilde x)^2 \rangle} = \hbar/\sqrt {12 mk_BT}, \eqno
(41)$$
\ni
i.e., inversely proportional to the square root of both temperature and mass.
For a hydrogen atom at 100 K, this is $0.2 {\AA}$.

Figure 12a shows the average RMS delocalization perpendicular to the hyperplane
as
a function of the reaction coordinate for $H_2$ at $100 ~K$ and at
$300~ K$.  At $100~ K$
the delocalization suddenly increases, in the space of one hyperplane, to a
much
higher value in the barrier region.  This delocalization is the source of the
flat-topped free energy barriers at low temperatures and is a clear signature
of tunneling.

The temperature where tunneling becomes important can be estimated from the
curvature of the potential barrier.
Consider a chain with its centroid atop a one dimensional parabolic barrier.
The images will spread away from the barrier top as far as the spring constant
allows.
The barrier top is thus not as unfavorable for a quantum atom as it is for a
classical atom.  The centroid density at the barrier top is
correspondingly increased, and any reaction that goes over the barrier will
proceed faster than expected.  This corresponds to tunneling.

The effective energy of a given configuration $x(\tau)$, which determines the
Boltzmann sampling of configurations
with $V(x)$ a parabolic barrier potential $V = -k x^2/2$ is
$$V_{eff}[x(\tau)] = \sum_{i=1}^P \left[ {{k_{spr}(x_i - x_{i-1})^2} \over 2} -
{1 \over P} {{k(x_i)^2} \over 2} \right]. \eqno (42)$$
Both the spring energy and the barrier potential scale as distance squared.
Thus, if the distances of all beads from the center are doubled ($x_i \to 2
x_i$),
both the spring energy and the barrier energy are quadrupled.  If the spring
constant $k_{spr}$ is small compared to the barrier sharpness $k$, $V_{eff}$ is
negative.  Then $V_{eff}$ decreases without bound as the beads move away from
the barrier, and the chain delocalizes to infinity in both directions.

In practice, when the springs are weak, the chain runs away from the barrier
top until the barrier starts curving back up.  The critical temperature, in
terms of the imaginary barrier frequency $\Omega \equiv k / \mu$, turns out to
be $k_B T_c = \hbar\Omega/(2\pi).^{12}$  This is a reasonable estimate of the
temperature
at the onset of tunneling.
Below $T_c$, the chain delocalization near the barrier is
expected to be significantly larger than in free space.

Fig. 12b shows the difference in RMS delocalization between the reactant and
transition state at a range of temperatures, for $H_2$ and $D_2$.
Significant increase in the delocalization sets in at
about $ \ 400 ~ K$ for hydrogen and at about $ \ 300 ~ K$ for deuterium,
as the temperature dependent spring constants in the FPI become weaker than
the curvature of the MEP potential barrier.
Using the curvature of the MEP potential barrier to estimated the
onset of increased delocalization and the equation for $T_c$ above, gives
$T_c$ = 400 K for $H_2$, and 290 K for $D_2$, in good agreement with the
observed behaviour.

The onset temperature of tunneling delocalization is well predicted from
a parabolic barrier reference, but the delocalization turns out to increases
smoothly as the temperature is lowered further, while the parabolic
barrier model predicts a very
sharp cutoff temperature below which the delocalization goes to infinity.
The gradual increase in the delocalization is due to the anharmonicity
of the MEP potential.  As the FPI images start probing regions where
the barrier becomes significantly non-parabolic
a balance is reached between the downward push from the potential gradient
and the upward pull from the springs connecting to images on the other side
of the barrier top. Thereby the tunneling delocalization saturates.
Fig. 13 shows the distribution of FPI images for hydrogen projected onto the
MEP in
typeical snapshots taken
from the statistical sampling at $100 ~ K $ and $300 ~ K$.
Even at $300~K$ where the tunneling effect is small,
the FPI images already probe regions
where anharmonic effects are clearly significant.

In $H_2$ at 300 K, the {\it change} in RMS delocalization between initial and
transition states (see figure 12b) is small.  However, the {\it actual}
delocalization is about $0.12 {\AA}$.  The delocalization goes to zero at high
temperature, but very reluctantly, as an inverse square root.  The quantum
system therefore is not expected to become truly classical except at extremely
high temperatures.

However, the quantum system at moderate temperatures such as $300~ K$ behaves
quasi-classically:  the delocalization remains roughly constant, and the
distribution of images remains roughly spherical.  The system's statistics
correspond roughly to those of a classical system at the same temperature, but
whose potential energy surface has been shifted to account for the zero-point
energy effects seen at the non-zero delocalization.  In other words, the
quantum effects at moderate temperatures, though non-zero, may simply show up
as an implicit shift in the potential energy, rather than as the fundamental
differences that are observed at low temperatures.

\vskip 1. true cm
\ni {\it 6.7 \  Torque contribution to the reversible work}

The rotational contribution to the reversible work turns out to be small
for this system.
Figure 14 shows the accumulated free energy difference due to rotation of the
hyperplanes as a function of the reaction coordinate.
While the curvature of the path is quite large, the torque with respect to the
MEP is small and the net effect on the free energy is small.
This is unlike the Eckart barrier plus harmonic oscillator model system,
where the rotational contribution was found to be just as large as
the translational contribution when using the MEP as reaction path.$^{10}$
There, tunneling caused the average density to move off the MEP
and thereby create torque on the hyperplane in a region of large path
curvature.  By redefining the path to coincide roughly with the average
density, the torque on the hyperplane and thereby the rotational
contribution to the free energy was minimized.  The net free energy barrier,
however, was the same.

While the rotational contribution is small here, it is statistically
significant,
and of opposite sign in the classical and quantum case (fig. 14).  Quantum
delocalization effects result in a shift of
the average density of the system with respect to the MEP,
which presumably
changes the direction of the torque and thereby the sign of the
rotational contribution to the free energy.

\vskip 1. true cm
\ni {\it  6.8 \ Activation energy for adsorption and desorption}

Fig. 15 shows an Arrhenius plot for the calculated free energy barriers
to dissociative adsorption.
The classical statistical sampling gives a straight line corresponding
to a constant activation energy of $0.73 ~ eV$.
Both $H_2$ and $D_2$ are observed to be classical at $600 ~K$ but
below $400~K$ the quantum results deviate from the classical results.
The quantum statistical
sampling gives significantly lower free energy barriers below $400~K$,
and a correspondingly lower activation energy of $0.38 ~ eV$ for $H_2$ and
$0.45 ~ eV$ for $D_2$.  A near Arrhenius behaviour is observed in the
quantum regime as well as in the classical regime;
the calculated points fall close to a
straight line fit.
The lowering of the activation energy is mainly
due to tunneling, as illustrated by the delocalization perpendicular
to the hyperplane shown in fig. 12b.
The tunneling isotope effect is $(0.73-0.38)/(0.73-0.45) \ =\  1.25$.

The experimental data of Campbell and coworkers is also shown
in fig. 15.  The measured activation energy agrees well with the
calculated results, but the measured sticking probability is larger
by a factor of $40$, indicating the entropy barrier is too large
in the simulation and the model potential should be adjusted to have
a wider reaction channel at the transition state.
If we assume the in-plane vibrations are harmonic, this can be accomplished,
for example, by reducing all five frequencies by a factor of
$1 / 40^{1/5} \approx 1/2$.
This takes into account only the five non-reactive hydrogen vibrations
(such as symmetric translation in the groove of the (110) surface, symmetric
and antisymmetric vibrations perpendicular to the groove, and antisymmetric
vertical motion)
and neglects modifications in the copper vibrations.  Since the shape
of our model potential surface is
quite arbitrarily derived from the LEPS potential form with no direct
experimental
input determining the width of the reaction channel at the transition state,
such a modification is not unreasonable.

Fig. 16 shows an Arrhenius plot for the reverse reaction, recombinative
desorption.
The classical statistical sampling gives Arrhenius type behaviour
with the calculated points falling near a straight line corresponding
to a constant activation energy of $0.55 ~ eV$.
The quantum statistical
sampling gives significantly lower free energy barriers below $400~K$,
and a correspondingly lower effective activation energy.
The calculated points do not fall on a straight line,
as is typical for decay of metastable states, and the effective activation
energy
is therefore somewhat temperature dependent.
At $300 ~ K $ the calculated activation energy is $0.49 ~ eV$ for $H_2$.
An experimental activation energy has been reported by Anger, Winkler, and
Rendulic$^{34}$ as $0.43\ eV$ in this temperature range and
in the zero coverage limit.
The calculated and measured activation energies are therefore in quite good
agreement.
The activation energy at $100~ K$ is calculated to be significantly lower, down
to $0.36~ eV$.  Again, tunnelling through the barrier has a strong effect
at low temperature.

\vskip 1. true cm
\ni {\it 6.8 \ Classical dynamical corrections}

The transmission coefficient, $\kappa$, can be
calculated by molecular dynamics.$^{8,16}$
An ensemble of points is found at the dividing surface whose momenta are
Boltzmann-distributed and initially forward-directed.
The ensemble is propagated forward and backward in time until all its members
leave the vicinity of the transition state.  The fraction of these trajectories
which actually started in R and finish in P is counted.
We assume that once the system leaves the vicinity of the transition state, it
will remain either a reactant or a product for a much longer time than it spent
near the transition state.  If, however, it spends a long time in the
intermediate region, dynamical corrections become problematic.

Classical dynamical correction calculations were carried out for $H_2$ and
$D_2$ at $600 ~K$, where classical statistical sampling gave the same TST
rate estimate as quantum statistical sampling.
An ensemble of 144 points was taken at the transition state hyperplane; each of
these was assigned a Boltzmann velocity and each trajectory was followed
forward and backward in time until it left the transition state region.  TST
assumes that each crossing point will finish as products when propagated
forward and as reactants when propagated backward.  If, for a given crossing
point, either of these assumptions is untrue, it does not count toward the
rate.  The transmission coefficient was found to be $\kappa = 0.54 \pm 0.08$
for $H_2$ and $\kappa = 0.56 \pm 0.08$ for $D_2$.  No kinetic isotope effect
was
observed for this system.

A transmission coefficient of $0.5$ has the same effect on the rate as does a
free energy increase of $-k_B T ln \kappa = 0.03 ~eV$ at $600~ K$.  This is of
the
same magnitude as the error associated with calculation of the free energy.
More
careful estimations of $\kappa$ are therefore not called for unless the free
energy difference is also calculated significantly more accurately.

None of the trajectories lingered in the transition region for more than a
picosecond.
The quantum dynamics are expected to be similar to
the classical dynamics at this high temperature,
and the quantum transmission coefficient
will, therefore, likely be within an order of
magnitude of the classical
transmission coefficient,
corresponding to a few hundredths of
an $eV$.
The free energy difference would then provide a good estimate of the
sticking coefficient, even without dynamical corrections.  Furthermore,
since the
classical trajectories spent little time near the transition state, it may be
feasible to do a quantum-dynamical calculation
for the short time required to estimate corrections to the QTST estimate.
However, we expect those corrections to be small, and
conclude from the classical dynamical corrections
that TST is an excellent approximation for treating the
$H_2$ dissociative adsorption as well as associative desorption at a $Cu(110) $
surface.  Other dissociative adsorption processes are likely to be also well
approximated by TST.

\vskip 1.0 true cm
\cl {\bf 7. Conclusions}

We have presented a calculation of the dissociative
sticking and thermal desorption
of hydrogen molecules using a statistical approach,
quantum transition state theory, where the two
hydrogen atoms were included fully quantum mechanically and a thermal
average was carried out over both the molecular and surface degrees of freedom.
For the model potential surface used here,
the system crosses over from classical to quantum behavior at about $T = 400 ~
K$ for $H_2$ and $T = 300 ~ K$ for $D_2$.  In the quantum regime, the
activation
energy is significantly reduced, the free energy barriers are flat-topped, and
the
delocalization is significantly greater at the barrier than in the reactant
state,
indicating that tunneling becomes the dominant transition mechanism.

A practical method for evaluating the required free energy barriers in
classical and quantum systems has also been given.
This method is based on a reversible work evaluation and can be applied
with relative ease to systems of high dimensionality.
Another advantage of this method
is the identification of an optimal, hyperplanar dividing surface
for the TST rate estimate in the course of the reversible work calculation.

\vfill\eject

\cl {\bf APPENDIX}

In this appendix we show that
the contributions to the free energy difference between
two hyperplanes due to the in plane forces vanishes
in the quantum mechanical
case, just as in the classical case.
We start with the rotational component of eqn. (25)
$$dA_{rot}(s)=\left\langle
\int_0^{\beta \hbar} {{d\tau } \over {\beta \hbar}}
{\bf \nabla} V[{\bf r}(\tau )] \cdot
{\bf U}^\prime_s {\bf U}^T_s
\left({\bf r}(\tau )-{\bf \Gamma}_s \right)
\right\rangle _s \ ds.\eqno (A1)$$

\ni
With the transformation
to local coordinates, {\bf z}, given by eqn. (12),
this expression may be rewritten in terms of the rotated coordinates ${\bf z}$
as
$$dA_{rot}(s)=\left\langle
\int_0^{\beta \hbar} {{d\tau } \over {\beta \hbar}}
{{\partial V[{\bf z}(\tau )]}\over{\partial {\bf z}}} \cdot
{\bf U}^T_s {\bf U}^\prime_s
{\bf z(\tau)}
\right\rangle _s \ ds \eqno (A2)$$
where the average
is written in terms of ${\bf z}$ as
$$\left\langle \cdots  \right\rangle _s =
{ {\int D{\bf z}(\tau ) e^{-S/ \hbar} \cdots \delta \left[ ( {\bf \tilde
z}_0)_1 \right] }
\over
{\int D{\bf z}(\tau ) e^{-S/ \hbar} \delta \left[ ( {\bf \tilde z}_0)_1 \right]
} }. \eqno (A3)$$

Consider the following average where $i\ne 1$ and $j\ne i$
$$I_{ij}=\left\langle
\int_0^{\beta \hbar} {{d\tau } \over {\beta \hbar}}
{{\partial V[{\bf z}(\tau )]}\over{\partial z_i }}
z_j(\tau)
\right\rangle _s .\eqno (A4)$$

In terms of a path integral, this becomes
$$I_{ij}= C  \int D{\bf z}(\tau ) e^{-S/ \hbar}
\int_0^{\beta \hbar} {{d\tau } \over {\beta \hbar}}
{{\partial V[{\bf z}(\tau )]}\over{\partial z_i }}
z_j(\tau)
\delta \left[ ( {\bf \tilde z}_0)_1 \right] . \eqno (A5)$$
where C is a normalization constant.

Using the following relation
$${{\partial S}\over{\partial ({\bf \tilde z}_0)_i}} =
\int_0^{\beta \hbar} d\tau
{{\partial V({\bf z}(\tau))}\over{\partial z_i}}$$
we may write
$$I_{ij}={{-C}\over \beta}
\int D{\bf z}(\tau )
{{\partial }\over{\partial ({\bf \tilde z}_0)_i}}
\left [
e^{-S/ \hbar} z_j(\tau)
\delta \left[ ( {\bf \tilde z}_0)_1 \right]
\right ] . \eqno (A6)$$

$I_{ij}$ can be shown to be zero in the following way.
First introduce
$$G \equiv \left [
e^{-S/ \hbar} z_j(\tau)
\delta \left[ ( {\bf \tilde z}_0)_1 \right]
\right ] \eqno (A7)$$
so that $I_{ij}$ can be written as
$$I_{ij}={{-C}\over \beta}
\int D{\bf z}(\tau )
{{\partial }\over{\partial ({\bf \tilde z}_0)_i}}
G \eqno (A8)$$
and
$$
G \rightarrow 0
\; as \;
({\bf \tilde z}_0)_i \rightarrow \pm \infty. \eqno (A9)$$

By transforming to Fourier variables
$${\bf \tilde z}_k = {1 \over {\hbar \beta}} \int_0^{\hbar \beta} d\tau
e^{-i\Omega_k\tau}{\bf \tilde z}(\tau) \eqno (A10)$$
where
$$ \Omega_k={{2 \pi k}\over {\hbar \beta}}  \eqno (A11)$$

\ni
eqn. (A6) becomes
$$I_{ij}={{-C}\over \beta}
J \left[ \prod_{(k,l) \ne (0,i)} \int d({\bf \tilde z}_k)_l \right]
\int d({\bf \tilde z}_0)_i
{{\partial }\over{\partial ({\bf \tilde z}_0)_i}}
G \eqno (A12)$$
where J is a constant Jacobian for the transformation
$$\int D{\bf z}(\tau) = J\prod_{(k,l)} \int d({\bf \tilde z}_k)_l .\eqno
(A13)$$

We next do the $({\bf \tilde z}_0)_i$ integration to give
$$I_{ij}={{-C}\over \beta}
J \left[ \prod_{(k,l) \ne (0,i)} \int d({\bf \tilde z}_k)_l \right]
\left(G(+\infty)-G(-\infty)\right) \ ds \eqno (A14)$$
which verifies
$$I_{ij}=0 . $$

Using this result and the fact that $ {\bf U}^T_s {\bf U}^\prime_s $ is
antisymmetric,
gives
$$dA_{rot}(s)\ =\ \left\langle
\int_0^{\beta \hbar} {{d\tau } \over {\beta \hbar}}
{{\partial V[{\bf z}(\tau )]}\over{\partial z_1}}
\left [
{\bf U}^T_s {\bf U}^\prime_s
{\bf z(\tau)}
\right ]_1
\right\rangle _s \ ds  \eqno (A15)$$
which can be rewritten as
$$dA_{rot}(s)\ =\  - \left\langle
\int_0^{\beta \hbar} {{d\tau } \over {\beta \hbar}}
\left [
{{\partial V[{\bf r}(\tau )]}\over{\partial {\bf r}}} \cdot {\bf U}_s
\right ]_1
\left [
{\bf U}^{\prime T}_s
\left({\bf r}(\tau )-{\bf \Gamma}_s \right)
\right ]_1
\right\rangle _s \eqno (A16)$$
and with $\left[ {\bf U}_s \right]_{i1} = \left[ {\bf n}_s \right]_i$,
we recover our final result
$$dA_{rot}(s)\ =\  - \left\langle
\int_0^{\beta \hbar} {{d\tau } \over {\beta \hbar}}
\left({{\partial V[{\bf r}(\tau )]}\over{\partial {\bf r}}} \cdot {\bf n}_s
\right)\ \left(
{\bf n}^\prime_s \cdot
\left({\bf r}(\tau )-{\bf \Gamma}_s \right) \right)
\right\rangle _s  \ ds \eqno (A17)$$
which is equivalent to eqn (27) derived for the classical case.

\vskip 2 true cm

\ni
{\bf ACKNOWLEDGEMENTS}

This work is supported by the Division of Chemical Sciences, Office of
Basic Energy Sciences, U.S. Department of Energy under
grant No. DE-FG06-91ER14224 (G.M. and H.J.)
and under Contract No.
DE-AC06-76RLO 1830 with Battelle Memorial Institute which
operates the Pacific Northwest Laboratory (G.S.).
G.M. acknowledges support from the Hertz foundation
and ARCS foundation.


\cl {\bf  \ \ References}
\vskip 1 true cm




\item { 1.} {See contributions by J. Harris, A. E. DePristo, S. Holloway and
           B. E. Hayden; in {\it Dynamics of Gas-Surface Interactions},
          edited by  C. T. Rettner and M. N. R. Ashfold (The Royal Society of
          Chemistry, Cambridge, 1991).}

\item { 2.} {B. Jackson and H. Metiu, {\it J. Chem. Phys.},  {\bf 86}, 1026
(1986);
C. -M. Chiang and B. Jackson, {\it J. Chem. Phys.}, {\bf 87}, 5497 (1987);
M. R. Hand and S. Holloway, {\it J. Chem. Phys.}, {\bf 91}, 7209 (1989);
J. Harris, {\it Surf. Sci.}, {\bf 221}, 335 (1989);
J. S. Sheng and J. Z. H. Zhang, {\it J. Chem. Phys.}, {\bf 99}, 1373 (1993).}

\item { 3.} {C. Engdahl and U. Nielsen, {\it J. Chem. Phys.}, {\bf 98}, 4223
(1993).}

\item { 4.} {U. Nielsen, D. Halstead, S. Holloway and J. K Norskov,
          {\it J. Chem. Phys.},     {\bf 93}, 2879 (1990).}

\item { 5.} {A. Gr\"uneich, A. J. Cruz and B. Jackson, {\it J. Chem. Phys.},
{\bf 98}, 5800 (1993).}

\item { 6.} {R. P. Feynman and A. R. Hibbs, {\it Quantum Mechanics and Path
           Integrals}, (McGraw Hill, New York, 1965). }

\item { 7.} {J. M. Campbell, M. E. Domagala and C. T. Campbell, {\it J. Vac.
Sci.
Technol.}, {\bf A9 (3)}, 1693 (1991); \
J. M. Campbell and C. T. Campbell, {\it Surf. Sci.},
{\bf 259}, 1  (1991).}

\item { 8.} {C. H. Bennett, in
 {\it Diffusion in Solids: Recent Developments} edited by J.J. Burton
and A.S. Nowick (Academic, New York, 1975).}

\item { 9.} {A. F. Voter,
 {\it J. Chem. Phys.}, {\bf 82}, 1890 (1985).}

\item {10.} {G. Schenter, G. Mills, and H. J\'onsson,
 {\it J. Chem. Phys.}, in press.}

\item {11.} {G. Mills and H. J\'onsson,
{\it Phys. Rev. Letters},
{\bf 72}, 1124 (1994).}

\item { 12.} {(a) \ M. J. Gillan, {\it J. Phys. C: Solid State Phys.}  {\bf
20}, 3621  (1987); \ \ (b)\  M. J.
Gillan, {\it Phys. Rev. Lett.} {\bf 58}, 563 (1987);\  \ (c)\  M. J. Gillan,
{\it  Philosophical Magazine A}, {\bf 58}, 257 (1988).}

\item { 13.} {
  P. Pechukas in {\it Dynamics of Molecular Collisions, Part B, }
  ed. W.H. Miller (Plenum Press, NY, 1976), p. 269;
  P. Pechukas, {\it Ann. Rev. Phys. Chem.} {\bf 32}, 159 (1981)}

\item { 14.} {W.H. Miller, Acc. Chem. Res. 9 (1976) 306.

\item { 15.} {A.F. Voter and J.D. Doll, J. Chem. Phys. 80 (1984) 5832; J.D.
Doll and A.F. Voter, Ann. Rev. Phys. Chem. 38 (1987) 413.}

\item {16.} {J. B. Anderson, {\it J. Chem. Phys.}, {\bf 58}, 4684 (1973).}

\item { 17.} {
E. Wigner, {\it J. Chem. Phys.}, {\bf 5}, 720 (1937);
J. C. Keck, {\it J. Chem. Phys.}, {\bf 32}, 1035 (1960); Adv. Chem.
Phys.}, {\bf 13}, 85 (1967); R. L. Jaffe, J. M. Henry and J. B. Anderson,
{\it J. Chem. Phys.}, {\bf 59}, 1128 (1973);
B. C. Garrett and D. G. Truhlar, {\it J. Phys. Chem.},
 {\bf 83}, 1052, 3058(E) (1979); {\bf 87}, 4553(E) (1983).}

\item { 18.} {E. Pollak and P. Pechukas, {\it J. Chem. Phys.} {\bf 69}, 1218
(1978)}

\item { 19.} {W. H. Miller,
 {\it J. Chem. Phys.}, {\bf 61}, 1823 (1974)}

\item { 20.} {B. C. Garrett, D. G. Truhlar, R. S. Grev, A. W. Magnuson,
 {\it J. Phys. Chem.}, {\bf 84}, 1730 (1980)}

\item { 21.} {J. W. Tromp and W. H. Miller,
 {\it J. Phys. Chem.}, {\bf 90}, 3482 (1986)}

\item { 22.} {G. A. Voth, D. Chandler, and W. H. Miller, {\it J. Chem. Phys.},
 {\bf 91}, 7749 (1989); G. A. Voth, {\it Chem. Phys. Lett.}, {\bf 170},
289 (1990); \ {\it J. Phys. Chem.}, {\bf 97}, 8365 (1993).}

\item { 23.} {A. A. Stuchebrukhov,
 {\it J. Chem. Phys.}, {\bf 95}, 4258 (1991).}

\item { 24.} {M. Messina, G. K. Schenter and B. C. Garrett,
 {\it J. Chem. Phys.}, {\bf 98}, 8525 (1993).}

\item { 25.} {M. Messina, G. K. Schenter and B. C. Garrett,
 {\it  J. Chem. Phys.}, {\bf 99}, 8644 (1993).}

\item { 26.} {F. J. McLafferty and P. Pechukas,
 {\it Chem. Phys. Letters}, {\bf 27}, 511 (1974).}

\item { 27.} {N. F. Hansen and H. C. Andersen,
 {\it (Submitted to J. Chem. Phys.)}}

\item { 28.} {
J. A. Barker, {\it J. Chem. Phys.}, {\bf 70}, 2914 (1979);
D. Chandler and P. G. Wolynes, {\it J. Chem. Phys.}, {\bf 74},
4078 (1981);
M. Parrinello, A. Rahman, {\it J. Chem. Phys.},   {\bf 80}, 860 (1984).}

\item { 29.} {T. R. Mattsson, U. Engberg, and G. Wahnstr\"om,
 {\it Phys. Rev. Letters}, {\bf 71} 2615 (1993).}

\item { 30.} {Y-C. Sun and G. A. Voth, {\it J. Chem. Phys.} {\bf 98}, 7451
(1993)}

\item { 31.} {H. C. Andersen, {\it J. Chem. Phys.}, {\bf 72}, 2384 (1980).}

\item { 32.} {W. H. Miller, S. D. Schwartz, and J. W. Tromp,
 {\it J. Chem. Phys.}, {\bf 79}, 4889 (1983).}

\item {33.} {C. T. Rettner, H. A. Michelsen and D. J. Auerbach,
            {\it Transactions of the Faraday Soc.}, (in press)}

\item {34.} {G. Anger, A. Winkler and K. D. Rendulic, {\it Surf. Sci.},  {\bf
220}, 1   (1989).}

\item {  35.} {M. Karikorpi, S. Holloway, N. Henriksen and J. K. Norskov,
 {\it Surf. Sci.}, {\bf 179}, L41 (1987)}

\item {  36.} {B. Hammer, K. W. Jacobsen and J. K. Norskov,
 {\it Phys. Rev. Letters}, {\bf 69}, 1971 (1992)}

\item { 37.} {Daw and Baskes,
 {\it Phys. Rev. B.} {\bf 29}, 6443 (1984)}

\item { 38.} {C.-Y. Lee and A. E. DePristo,
 {\it J. Chem. Phys.} {\bf 85}, 4161 (1986).}

\item { 39.} {M. R. Hand and J. Harris, {\it J. Chem. Phys.}, {\bf 92}, 7610
(1990).}

\item { 40.} {H. J\'onsson and G. Mills,
 {\it (Submitted to J. Chem. Phys.)}}

\item { 41.} {The analogous figure in ref. 9 is different because
the potential energy curve shown there
is for a frozen substrate and does not correspond to the MEP.
Also, there are slight differences in the interaction potential
and the coupling to the thermal bath between the present calculations
and ref. 9 leading to small differences in
the calculated results.

}


\cl {\bf Table 1: \ \ Properties of the empirical interaction potential}

\vskip 0.8 true cm

\settabs 3 \columns

\+ Characteristic & Experimental & Fitted \cr

\ni \cl {$H_2$:}
\+ {Separation (\AA)} & 0.737 & 0.740 \cr
\+ {Binding energy (eV)} & -4.747 & -4.7472 \cr
\+ {Vibration (eV/\AA$^2$)} & 24.50 & 25.43 \cr
\vskip 0.3 true cm

\ni \cl {$H-Cu(110)$:}

\settabs 3\columns
\+ {Binding energy (eV)$^{7,34}$} & -2.3 & -2.29 \cr
\+ {Vibration (eV/\AA$^2$)$^{38}$} & 5.5 & 5.0 \cr

\vskip 0.3 true cm
\ni \cl {Long Bridge Saddle Point:}

\settabs 3\columns
\+ {Saddle point energy (eV)} & {} & +0.688 \cr
\+ {H-H separation (\AA)} & {} & { 1.3385} \cr
\+ {H-Cu height (\AA)} & {} & { 0.8802} \cr


\vfill\eject
\vskip 2 true cm

\cl {\bf Table 2: \ \ Parameters of the interaction potential (units of eV and
\AA)}

\vskip 0.5 true cm

\settabs 4\columns
\+ Parameter  &  Cu         &     H        &   Cu-H    \cr
\+ $D_A$      &  2862       &    79.50     &  86.15    \cr
\+ $\alpha_A$ &  3.512      &    2.480     &  4.211    \cr
\+ $D_B$      & -109.1      &   -107.6     &  1.536 E+04  \cr
\+ $\alpha_B$ &  1.756      &    2.999     &  6.076    \cr
\+ $S$        &  0.273      &    2.144     &  \cr
\+ $\beta_A$  &  3.691      &    3.777     &  \cr
\+ $\beta_B$  &  7.381      &    0         &  \cr
\+ $\eta$     &  6          &    0         &  \cr
\+ $\gamma$   &  512        &    0         &  \cr
\+ $f_1$      & -112.9      &   -81.75     &  \cr
\+ $f_2$      &  8510       &    838.7     &  \cr
\+ $f_3$      & -2.617 E+05 &   -3953      &  \cr
\+ $f_4$      &  4.780 E+06 &    8768      &  \cr
\+ $f_5$      & -5.234 E+07 &   -6599      &  \cr
\+ $f_6$      &  3.391 E+08 &              &  \cr
\+ $f_7$      & -1.201 E+09 &              &  \cr
\+ $f_8$      &  1.796 E+09 &              &  \cr


\cl {\bf  \ \ Figure captions}
\vskip 1 true cm

\ni
{\bf Fig. 1}. \ \
(a) \ Reaction path $\Gamma $
parametrized with reaction coordinate $s$.
For each point along the path, a hyperplane
intersecting the  path at that point and having
a normal $\bf n_s$
is defined, $Z = {\bf n_s} \cdot ({\bf r}
- {\bf \Gamma_s}) = 0$.
The force acting on the system is
averaged over a statistical sample of configurations
where the system is
constrained to lie in the hyperplane.

(b)
The hyperplane is gradually moved from the reactant region towards
products by varying the reaction coordinate, $s$.
The activation free energy for the transition
is obtained by calculating the reversible work involved in moving and
rotating the hyperplane to an intermediate position
chosen to be the dividing surface between reactant and product regions.
The maximum free energy barrier is obtained
where the force and torque
on the hyperplane vanish. This is equivalent to finding the
optimal hyperplanar dividing surface in classical variational transition state
theory.

\vskip 1 true cm

\ni
{\bf Fig. 2}. \ \
The system confined to a hyperplane experiences a gradual change in
the potential energy as the hyperplane is moved along the reaction
path.  \ \ (a) A schematic contour plot of a potential energy surface
with the reaction path shown as a thick solid line.  Two different
locations of the hyperplane are indicated along the progression from reactants
to
products. They intersect the reaction path at ${\bf \Gamma}_s$
and ${\bf \Gamma}_{s+ds}$ and their orientation is given by the
normal vectors ${\bf n}_s$ and ${\bf n}_{s+ds}$ as shown in the inset.
\ \ (b) Potential energy curves corresponding
to the two hyperplanes.  This illustrates the ramping of the potential
experienced by the system confined to stay within the hyperplane,
as the hyperplane is moved from one location to another.

\vskip 1 true cm

\ni
{\bf Fig. 3}. \ \
In the quantum mechanical evaluation of the free energy barrier
the hyperplane constraint is applied to the centroids, ${\bf \tilde r}_0$,
of the Feynman path integral representations of the quantum particles
in the system,
instead of the classical atom coordinates (compare with fig. 1a).
The force acting on the FPI images is evaluated and
averaged over a statistical sample of configurations,
with the centroids of the FPI chains
constrained to lie in the hyperplane.

\vskip 1 true cm

\ni
{\bf Fig. 4}. \ \
The minimum energy path between the molecular hydrogen outside the
surface and the dissociated hydrogen chemisorbed on adjacent sites
on the Cu(110) surface. The Nudged Elastic Band method was used
to find the optimal path.
The initial state, shown in the upper left panel, was
arbitrarily chosen to have the molecule oriented with its axis normal to
the surface plane and located directly above a surface atom.
The final state, shown in the upper right panel,
has the H atoms sitting in fourfold sites
separated by a long bridge.
Along the minimum energy path, shown in the lower panel,
the molecule shifts over to the the long bridge
site and turns to make the axis parallel to the surface.  The path
from then on is symmetric and involves motion perpendicular to the
surface and perpendicular to the long bridge.  The molecule
breaks up very close to the surface.

\vskip 1 true cm

\ni
{\bf Fig. 5}. \ \
Potential energy contours for the dissociation of $H_2$
on $Cu(110)$ over a long bridge site.
Because of the high symmetry of the MEP shown in fig. 4, a two
dimensional cut through the six dimensional surface effectively
describes the energetics.
The molecule is coming down parallel to the surface and is symmetrically
arranged over a long bridge site;
$z$ is the height above the surface and $R$ the intramolecular
distance.

\vskip 1 true cm

\ni
{\bf Fig. 6}. \ \
Scatter plot of the orientation of the $H_2$ molecule in the statistical
sampling of the classical system at $T = 100 ~K$.
Theta is the polar angle, between $0$ and $\pi$.  Phi is the azimuthal angle,
between $-\pi$ and $\pi$.
\ \
(a) \  Points taken from the sampling of a hyperplane near the
reactant region.  The hyperplane constraint does not restrict the
rotation of the molecule.
\ \
(b) \  Points taken from the sampling of the transition state hyperplane,
illustrating the rotational confinement of the molecule due to the
$H_2-Cu$ interaction potential.

\vskip 1 true cm

\ni
{\bf Fig. 7}. \ \
Snapshots from the statistical averaging of the transition state
hyperplane at $T=100 ~K$.  Only one of the six centroid degrees of
freedom of the H atoms is constrained, so the system still has five degrees
of freedom to explore the potential energy surface.
In particular, the hyperplane constraint does not prevent
simultaneous movement of
the H atoms perpendicular to the bridge, as is evident from the
snapshot shown in the top panel.
However, the centroids
are not allowed to move closer together unless they simultaneously descend
toward the surface.  They are allowed to move apart only if they ascend higher
above the surface.  This hyperplane constraint prevents the atoms from either
chemisorbing on adjacent lattice sites, or desorbing.
The images of each of the FPI chains tend to separate
into two clusters illustrating the sliding down on opposite sides of
the potential barrier.  This is a signature of tunneling.

\vskip 1 true cm

\ni
{\bf Fig. 8}. \ \
The accumulated free energy difference with respect to the reactant state
as the hyperplane is moved along the reaction path in
a classical statistical sampling.  For reference the solid line
shows the potential energy along the minimum energy path.
The free energy barrier increases with temperature, indicating a
decrease in entropy in the transition state.

\vskip 1 true cm

\ni
{\bf Fig. 9}. \ \
Free energy curves obtained in classical statistical sampling
after freezing out selected degrees of freedom of the system.
\ \
(a) \ The lattice is made rigid to illustrate the effect of
confinement of the $H_2$ molecule as the transition state is
approached.  The entropy barrier becomes larger when the lattice
is rigid (compare with fig. 8), illustrating that relaxation in
the lattice relieves the confinement of the molecule in the
transition state.
\ \
(b) \ The dimer is held rigid on the reaction path while the lattice
degrees of freedom are sampled.  The molecule confines the lattice
vibrations significantly only at high temperatures,
$T \ >\  300 ~ K$.

\vskip 1 true cm

\ni
{\bf Fig. 10}. \ \
Quantum statistical sampling of the hyperplane along the reaction path
(quantum mechanical extension of the results in fig. 8)
showing a significantly lower free energy barrier at low temperature
as compared with the classical statistical sampling.
Already at $T \ =\  300 ~ K$ quantum effects have lowered the free energy
barrier by $ca.\  0.1 ~ eV$.  At $100 ~ K$ the barrier is flat-topped
due to large tunneling effects.
The free energy curve drops below the potential energy curve
early along the reaction path, well before tunneling is effecctive,
due to lowering of the zero point energy of the molecule.

\vskip 1 true cm

\ni
{\bf Fig. 11}. \ \
The accumulated free energy at $T \ = \ 100 ~ K$ for
a quantum statistical sampling of $D_2$ dissociative adsorption
The classical and quantum sampling of $H_2$ dissociative adsorption
from figs. 8 and 10 is shown for comparison.  The classical sampling
gives the same results for $D_2$ and $H_2$.
This shows that significant lowering of the free energy barrier is found
for $D_2$, but smaller than for $H_2$.

\vskip 1 true cm

\ni
{\bf Fig. 12}. \ \
(a)\
The RMS delocalization of $H_2$ perpendicular to the hyperplane as a
function of the reaction coordinate.  In a free $H$ atom the
delocalization is $0.2 ~ {\AA}$ at $100~K$. As the repulsive interaction
between the molecule and the surface sets in the delocalization
decreases.  Near the transition state, a dramatic increase
in the delocalization is observed at $100 ~ K$ due to tunneling.
In this segment of the reaction path, the free energy is nearly constant
(see fig. 11).
At $300 ~ K$, the delocalization is much weaker but still noticeable.
\ \
(b) \
The difference in the RMS delocalization perpendicular to the transition state
hyperplane and the delocalization of
free $H_2$ molecule and $D_2$ molecule as a function of inverse
temperature.
Straight lines fitted to the larger values are shown only to guide the eye.
Significant increase in the delocalization sets in at
about $ \ 400 ~ K$ for hydrogen and at about $ \ 300 ~ K$ for deuterium,
as the temperature dependent spring constants in the FPI become weaker than
the curvature of the MEP potential barrier.

\vskip 1 true cm

\ni
{\bf Fig. 13}. \ \
Distribution of FPI images for hydrogen projected onto the MEP in snapshots
taken
from the statistical sampling at $100 ~ K $ and $300 ~ K$.  The dashed line
shows a parabolic barrier with curvature matching that of
the MEP potential curve.  Even at $300~K$ the FPI images probe regions
where anharmonic effects are significant.  The vertical position of the
images is arbitrary.
At $300~K$ the energy of the system is high, several eV over the barrier,
and the distribution of images does not reflect the shape of the barrier.
At $100~K$ the energy of the system is closer to the barrier energy and
the effect of the barrier on the distribution is clear.

\vskip 1 true cm

\ni
{\bf Fig. 14}. \ \
Rotational contribution to the free energy
as a function of reaction coordinate at $T = 100 ~K$.
The contribution is negligible as compared with the translational
contribution in this system
when the the MEP is chosen to be the reaction path,
because the torque acting on the hyperplanes turns out to be very small.
A small but statistically significant contribution is obtained where the
path curvature is largest, near the transition state.  Quantum
delocalization effects result in a shift of
the average density of the system with respect to the MEP which
presumably
changes the direction of the torque and thereby the sign of the
rotational contribution to the free energy.

\vskip 1 true cm

\ni
{\bf Fig. 15}. \ \
An Arrhenius plot for $H_2$ and $D_2$ dissociative adsorption.
The calculated points are for $100~ K$, $150~ K$, $200 ~K$, $300~ K$, $400~ K$,
and $600 ~K$.
The classical statistical sampling gives a straight line corresponding
to a constant activation energy of $0.73 ~ eV$.  The quantum statistical
sampling gives significantly lower free energy barriers below $400~K$
and a correspondingly lower activation energy of $0.38 ~ eV$ for $H_2$
(slope of large dashed line)
and
$0.45 ~ eV$ for $D_2$ (slope of small dashed line).
The lowering of the activation energy is mainly
due to tunneling.
The experimental data of Campbell and coworkers$^{7}$ is shown with
triangles.  The measured activation energy agrees well with the
calculated results, but the measured sticking probability is larger
by a factor of $40$, indicating the entropy barrier is too large
in the simulation and the model potential should be adjusted to have
a wider reaction channel at the transition state.
A solid line is drawn through the calculated points only to guide the eye.

\vskip 1 true cm

\ni
{\bf Fig. 16}. \ \
An Arrhenius plot for $H_2$ and $D_2$ associative desorption.
The classical statistical sampling gives a straight line corresponding
to a constant activation energy of $0.55 ~ eV$.
The quantum statistical
sampling gives significantly lower free energy barriers below $400~K$,
and a correspondingly lower activation energy.
At $300 ~ K $, the calculated activation energy is $0.49 ~ eV$ for $H_2$ as
compared to the experimental value $0.43 ~ eV$ obtained by Anger et al.
(the solid line has a slope consistent with the experimental measurements
but its vertical position was simply chosen in such a way as to intersect
the origin, since an absolute value of the desorption rate was not reported).
A straight dashed line is drawn through the calculated points only to guide the
eye.

\bye